\documentclass[
 twocolumn, superscriptaddress, amsmath,amssymb,
 aps, prl
]{revtex4-1}

\usepackage{hyperref}
\usepackage{graphicx}
\usepackage{dcolumn}
\usepackage{bm}
\usepackage{xcolor}

\usepackage[mathlines]{lineno}

\usepackage{amsmath}

\begin{document}

\preprint{APS/123-QED}

\title{Nonlocal phase modulation of multimode, continuous-variable twin beams}

\author{Zhifan Zhou}
\affiliation{Joint Quantum Institute, National Institute of Standards and Technology and the University of Maryland, College Park, Maryland 20742, USA}

\author{Lu\'{i}s E.~E.~de Araujo}
\affiliation{Joint Quantum Institute, National Institute of Standards and Technology and the University of Maryland, College Park, Maryland 20742, USA}
\affiliation{Institute of Physics Gleb Wataghin, University of Campinas (UNICAMP), 13083-859 Campinas, S\~{a}o Paulo, Brazil}

\author{Matt DiMario}
\affiliation{Joint Quantum Institute, National Institute of Standards and Technology and the University of Maryland, College Park, Maryland 20742, USA}

\author{B. E. Anderson}
\affiliation{Department of Physics, American University, Washington DC 20016, USA}

\author{Jie Zhao}
\affiliation{Joint Quantum Institute, National Institute of Standards and Technology and the University of Maryland, College Park, Maryland 20742, USA}

\author{Kevin M. Jones}
\affiliation{Department of Physics, Williams College, Williamstown, Massachusetts 01267, USA}

\author{Paul D. Lett}
\email[Email: ]{lett@umd.edu}
\affiliation{Joint Quantum Institute, National Institute of Standards and Technology and the University of Maryland, College Park, Maryland 20742, USA}
\affiliation{Quantum Measurement Division, National Institute of Standards and Technology, Gaithersburg, Maryland 20899, USA}

\date{\today}

\begin{abstract}
We investigate experimentally the nonlocal phase modulation of multiple-frequency-mode, continuous-variable entangled twin beams. We use a pair of electro-optical phase modulators to modulate the entangled probe and conjugate light beams produced by four-wave mixing in hot Rb vapor. A single phase modulator in either one of the twin beams reduces the two-mode squeezing signal, and we find that the modulations interfere nonlocally to modify the beam correlations. The nonlocal modulation of the beams can produce quantum correlations among frequency modes of the multimode fields.
\end{abstract}

\maketitle

Multimode entanglement has attracted great interest in quantum computing~\cite{Furusawa2013, Anderson2019, Furusawa2019}, quantum metrology~\cite{guo2020distributed, malia2022distributed, Lucas2022}, and quantum information processing in general~\cite{Zeilinger2020}. Examples include Gaussian Boson sampling~\cite{GBS2017, GBS2019, GBS2020} and distributed quantum sensing~\cite{guo2020distributed, malia2022distributed, Lucas2022}, which rely on linear mixing and splitting of squeezed light sources. The technical challenge in these applications increases as the number of modes increases. Multimode entanglement in the frequency domain, such as in quantum optical frequency combs~\cite{Treps2013, Treps2014, Treps2017, Treps2020}, provides scalable and compact sources for those applications.

Twin beams play an important role in continuous-variable (CV) quantum information processing~\cite{braunstein2005} by enabling deterministic generation, manipulation, and detection of entangled light. The generation and control of CV entangled states of light find many  applications in quantum erasing~\cite{filip2003,andersen2004}, quantum steering~\cite{handchen2012}, quantum imaging~\cite{boyer2008}, quantum key distribution protocols~\cite{filip2005}, and cluster-state or measurement-based quantum computing~\cite{zhu2021}, among others. Recent work has suggested using an electro-optical phase modulator (EOM) for generating high-dimensional entangled states~\cite{zhu2021}.

Entanglement between particles persists even when they are spatially separated. This nonlocal character of quantum theory has been observed in many effects such as dispersion cancellation~\cite{franson1992,baek2009}, quantum erasing~\cite{ma2012}, aberration cancelation~\cite{black2019}, and phase modulation~\cite{harris2008,sensarn2009,seshadri2022}. In nonlocal phase modulation, two distant EOMs, each operating on one of a pair of entangled fields, act cumulatively to determine the apparent modulation. In the discrete-variable (DV) framework, nonlocal modulation is observed in the correlations between spatially-separated entangled twin photons~\cite{sensarn2009}.

In this Letter, we experimentally investigate the effect of nonlocal phase modulation with EOMs to mix frequency modes of CV two-mode squeezed beams. In the CV framework explored here, we are concerned with amplitude and phase correlations between quadratures of the twin fields. We restrict ourselves to a low modulation index so that we need not worry about mixing beyond the first sidebands. The EOM acts as a multi-port beamsplitter in frequency-mode space. While a simple beamsplitter will couple, say, the $X$ (amplitude) quadrature at the input to the $X$ quadratures in the output modes, a phase modulator such as we use here will couple the $X$ quadrature at one frequency to the $P$ (phase) quadratures in neighboring frequency bins. We describe our investigation into how an EOM affects the two-mode squeezing between twin beams and how the relative phase between a pair of EOMs, each acting on one of the twin beams, changes the correlations between neighboring frequency modes of the conjugate joint field quadratures.

To model the action of the EOM, we consider two-mode squeezed states produced in the double-lambda four-wave-mixing (4WM) scheme illustrated in Fig.~1. Probe and conjugate fields go through separate EOMs. Assuming the EOMs impart to the fields a periodical modulation at the same frequency, but allowing for different relative driving phases, we find the noise of the joint quadrature operator to be~\cite{noise,Suppl2023}
\begin{equation}
\begin{split}
     \langle \overline{X_{-}^2} \rangle  = &   (G^2 + g^2) (1-\eta) +\eta - 2 g G (1-\eta) \\
    & \times J_0\left(\sqrt{m_\text{p}^2 + m_\text{c}^2 + 2 m_\text{p} m_\text{c} \cos \phi} \right),
    \end{split}
\label{eq5}
\end{equation}
where $J_0$ is a zeroth order Bessel function, $G$ is the amplitude gain of the 4WM process, $g^2 = G^2 -1$; $\eta$ is the total loss for each of the probe and conjugate fields; $m_{\rm{p},\rm{c}}$ is the modulation index for the modulator acting on the probe (p) or conjugate (c) field; $\phi$ is the phase difference between the two EOMs; and the overbar represents a time average over one modulation period. If the modulation indices of the EOMs are equal ($m_\text{p} = m_\text{c} = m$), then
\begin{equation}
\begin{split}
    \langle \overline{X_{-}^2} \rangle = & (G^2 + g^2)(1-\eta) +\eta - \\ 
    & 2 g G (1-\eta) J_0 \left(m \sqrt{2 + 2 \cos \phi} \right).
\label{eq6}
\end{split}
\end{equation}
\noindent
If both EOMs are turned off ($m=0$), and there are no losses ($\eta = 0$), then $\langle \overline{X_{-}^2} \rangle = (G -g)^2$. Then for any $G^2>1$,  $\langle \overline{X_{-}^2} \rangle < 1$, so that the noise power is below shot noise and squeezing is observed. In deriving Eq.~\eqref{eq6}, we made no assumptions regarding the spatial separation of the probe and conjugate field modes. Equation~\eqref{eq6} implies that the maximum squeezing between the two fields is obtained when the two phase modulators are off. Turning the modulators on will, in general, reduce the degree of squeezing. For a high enough modulation index, squeezing may be eliminated as the quadrature noise will exceed the shot noise.

Three cases are particularly of interest: the EOMs are driven (in-phase) with a relative phase of $\phi = 0^{\circ}$ and (out-of-phase) with $\phi = 180^{\circ}$ and $\phi = 120^{\circ}$. When $\phi = 180^{\circ}$, Eq.~\eqref{eq6} clearly shows that the modulation imparted to one of the twin beams cancels the modulation experienced by the other twin beam; and two-mode squeezing is recovered at the same level as obtained with the EOMs off. Comparing Eqs.~\eqref{eq5} and \eqref{eq6}, we see that, when the two EOMs are driven in phase, they produce the same amount of squeezing as only one modulator operating at twice the modulation index ($m_\text{p} = 0$ and $m_\text{c} = 2 m$, or vice-versa). And for $\phi = 120^{\circ}$, Eq.~\eqref{eq6} predicts that the two EOMs should behave with respect to the two-mode squeezing signal as a single modulator driven at a modulation index of $m$. More generally, the effect of two phase modulators on the joint quadrature noise is similar to that of a single modulator operating at an effective modulation index of $\sqrt{m_\text{p}^2 + m_\text{c}^2 + 2 m_\text{p} m_\text{c} \cos \phi}$. In other words, with respect to the two-mode squeezing signal, the modulators act cumulatively to determine the effective modulation index, analogously to the DV case~\cite{harris2008,sensarn2009}. This cumulative effect is also nonlocal. That is, it is independent of the distance between the EOMs.

Figure~1 shows the experimental setup and a level diagram of the double-lambda 4WM scheme. The setup is similar to the one described in \cite{boyer2008}. A 12~mm-long Rb vapor cell is heated to 123~$^{\circ}$C.  A single pump beam with 700~mW of power and 650~$\mu$m beam diameter is detuned 0.8~GHz blue from the $^{85}$Rb $D_1$ line and is split equally into two beams: one beam goes through the vapor cell along with a probe seed beam to generate the local oscillator (LO) beams, and the other beam generates the two-mode squeezed vacuum states, as in \cite{boyer2008}. The LO probe seed is derived from the pump beam by double passing a small portion of the pump through a 1.5~GHz acoustic-optic modulator. Pump and probe intersect inside the cell at an angle of 7 mrad. A double-lambda 4WM process uses the $\chi^{(3)}$ nonlinearity of the Rb vapor to convert two pump photons into one probe photon and one conjugate photon.  (The squeezed signal beams are vacuum seeded.) The probe beam experiences a typical gain of 3. The non-degenerate probe and conjugate beams are at the same angle on opposite sides of the pump and in a two-mode squeezed state. Probe and conjugate beams pass through identical EOMs driven with a 200~kHz sine wave. The EOMs are driven synchronously by separate outputs of the same function generator, and their relative phase can be adjusted by the function generator. The modulated beams are sent separately to two balanced homodyne detectors, one for the probe and another for the conjugate.

In the homodyne detectors, the probe and conjugate beams are mixed with LO beams on 50/50 beamsplitters with fringe visibilities $>97 \,\%$. The relative phases $\theta_{p,c}$ between the LOs and the probe/conjugate fields are adjusted by mirrors mounted on piezoelectric transducers (PZT) in order to select the quadrature to be detected in each beam. The outputs of the homodyne detectors are directly measured with matched photodiodes with quantum efficiencies $>95 \, \%$. The path length from the vapor cell to the optical detectors for the probe and conjugate beams are approximately matched. Due to the different group velocities of the probe and conjugate beams in the atomic vapor \cite{boyer2007}, the two fields are optically delayed by amounts that differ by approximately 10~ns. To compensate for this delay, we add an electronic delay line after detection by adjusting the cable lengths. The photocurrents are amplified and then measured with a 1~GHz digital sampling oscilloscope. The measured time traces are digitally post processed to determine the power spectra and generalized quadrature noise powers discussed below.

\begin{figure}[htbp]
\centering
\includegraphics[width=8.5cm]{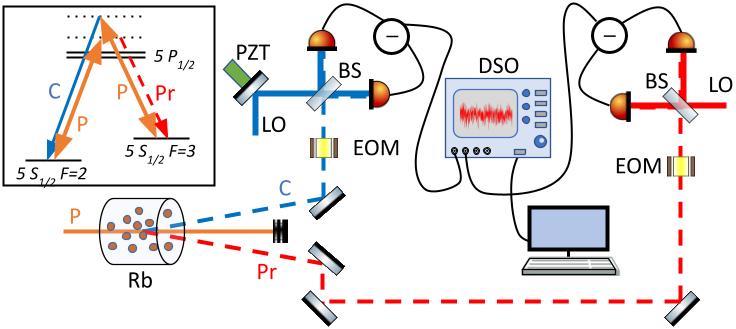}
\caption{Experimental setup and energy level diagram (inset) of the 4WM process in $^{85}$Rb. The pump beam (P) generates twin probe (Pr) and conjugate beams (C) in a two-mode squeezed state. BS are 50/50 nonpolarizing beamsplitters; EOMs are electro-optical phase modulators; PZT is a piezoelectric transducer; DSO is a 1~GHz digital sampling oscilloscope; and LO are local oscillator fields for the homodyne detection schemes.}
\label{fig1}
\end{figure}

The homodyne signal as a function of local oscillator phase gives a generalized quadrature: $\hat{X}_{i} \cos\theta_i + \hat{P}_{i} \sin\theta_i$. If we subtract the homodyne signals, we measure the noise power of the joint quadrature $\hat{X}_{\theta} = \hat{X}_{\theta_\text{p}} - \hat{X}_{\theta_\text{c}}$, where $\theta = \theta_\text{p} + \theta_\text{c}$. A typical noise spectrum as a function of the phase $\theta$ is shown in Fig.~2. Squeezing is observed when $\theta = 0$ (point I in the figure). A frequency-dependent squeezing spectrum is observed by locking the LO phase $\theta$ to point I by a noise locking technique~\cite{mckenzie2005}. (Locking the phase to point III, gives us the quadrature $\hat{P}_{\theta} = \hat{P}_{\theta_\text{p}} + \hat{P}_{\theta_\text{c}}$.) Because the temporal modulation imparted by the EOMs to the beams may disturb the locking signal, we pulse the driving signal from the function generator to the EOMs at 40 Hz. The driving pulses are square pulses with a width of 12.5 ms. The signals from the probe and conjugate homodyne detectors consisted of $10^6$ values sampled over 10 ms captured during the time the EOMs were on. The measurement window is intentionally smaller than the pulse width in order to avoid possible transient edge effects in the data. We used Welch's method~\cite{heinzel2002} to obtain the final power spectrum of the measured noise.

\begin{figure}[htbp]
\centering
\includegraphics[width=8cm]{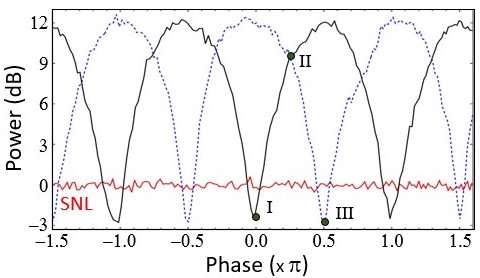}
\caption{Noise power of the sum (dashed blue line) and difference (solid black line) of the quadratures measured by the homodyne detectors as the phase $\theta$ is varied. In both cases, the noise is analysed at a frequency of 1~MHz. By locking the phase to points I ($\theta = 0$), II ($\theta = \pi/4$) or III ($\theta = \pi/2$), we can measure the joint quadratures \textit{XX}, \textit{XP} or \textit{PP}, respectively, of the twin beams.}
\label{fig2}
\end{figure}

Typical two mode squeezing spectra ($\langle \overline{X_{-}^2} \rangle$ vs.~frequency) taken with the LO phases locked at point I are shown in Fig. 3. When the EOMs are off (Fig.~3a), the squeezing spectrum extends over a bandwidth of approximately 15 MHz. Turning one EOM on (on either the probe or conjugate beam) with $m = 0.1 \pi$ reduces squeezing at all frequencies. At twice the modulation index, squeezing is eliminated, with the noise well above the shot noite. When both EOMs are on with the same modulation index, the effect on the entangled signal depends on their relative phase (Fig.~3b). When the EOMs are driven in phase, they act together to reduce the squeezing signal, producing a spectrum similar to that of a single EOM with twice the modulation index acting on only one of the beams (Fig.~3a). When the EOMs are driven $180^{\circ}$ out of phase, their effect on the two-mode squeezing cancels. With the EOMs driven at a relative phase of $120^{\circ}$, the squeezing spectrum is similar to that seen with only one EOM on. These results are in agreement with the predictions of our model. They are the CV analog of the nonlocal modulation effect reported in Ref.~\cite{sensarn2009} in the DV regime.

\begin{figure}[htbp]
\centering
\includegraphics[width=8.5cm]{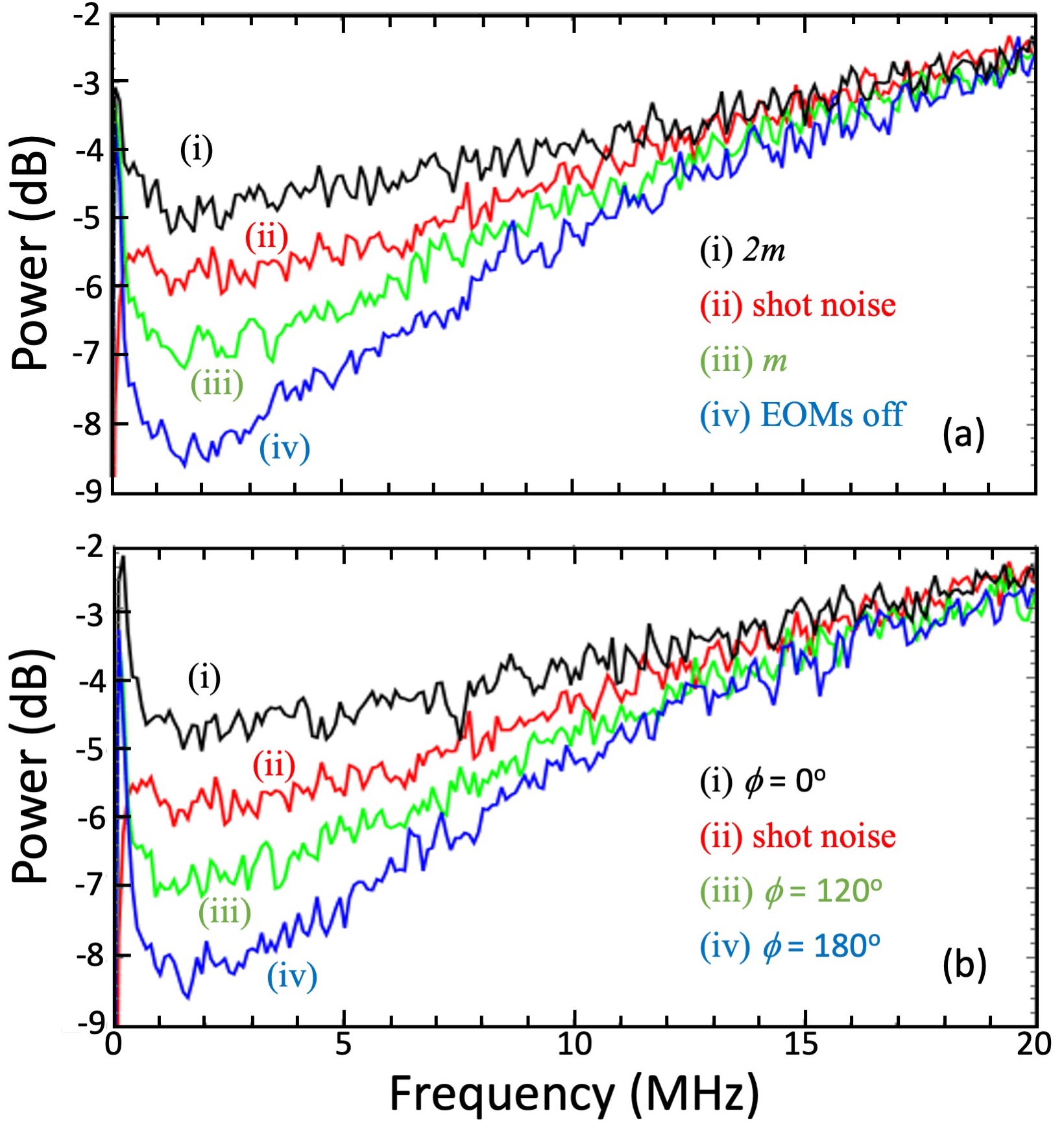}
\caption{(a) Squeezing spectra obtained with both EOMs turned off (blue line) and with one EOM on, but at different modulation indexes: $m$ (green line) and $2 m$ (black line); (b) Squeezing spectra for both EOMs running, at a modulation index $m$, in phase ($\phi = 0^{\circ}$) and out of phase ($\phi = 180^{\circ}$ and $\phi = 120^{\circ}$). In all cases, $m = 0.1 \pi$. Electronic noise was not subtracted and thus the shot  noise (red line) does not appear to be independent of frequency here~\cite{Suppl2023}.}
\label{fig3}
\end{figure}

Full characterization of the two-mode squeezed states requires determining the covariance matrix $C$ of the fields. In the ordered basis $(\hat{X}_\text{p},\hat{X}_\text{c},\hat{P}_\text{p},\hat{P}_\text{c})$:
\begin{equation}
C =
\begin{bmatrix}
C_{\textit{XX}} & C_{\textit{XP}} \\
(C_{\textit{XP}})^T & C_{\textit{PP}}
\end{bmatrix},
\end{equation}
where $C_{\textit{XX}}$, $C_{\textit{XP}}$, and $C_{\textit{PP}}$ are $2\times2$ matrices. The covariance matrix of the two-mode squeezed state is symmetric. $C_{\textit{XX}}$ and $C_{\textit{PP}}$ are associated with the amplitude \textit{XX} and phase \textit{PP} joint quadratures of the twin beams, respectively, while $C_{\textit{XP}}$ is the mutual correlation matrix between their \textit{X} and \textit{P} quadratures. When the EOMs are off, $C_{\textit{XP}} = 0$, so the covariance matrix is block diagonal. Turning on the phase modulators couples the \textit{X} and \textit{P} quadratures of the fields, and $C_{\textit{XP}} \neq 0$.

We can gain further insight into the characteristics of the nonlocal modulation of the EOMs by measuring the \textit{XP} quadrature of the covariance matrix for the twin beams. For that, we lock the joint quadrature phase to $\theta = \pi/4$ (point II in Fig.~2). In this way, we measure different quadratures ($X$ of the probe and $P$ of the conjugate) of the two beams. The locking scheme and how we recover the \textit{XP} covariance matrix from the acquired time traces are detailed in~\cite{Suppl2023}.

Figure 4a shows the measured \textit{XP} covariances. When the beams are not modulated, their \textit{X} and \textit{P} quadratures are not coupled. We also do not observe any correlations when the EOMs are operated at $180^{\circ}$ phase difference due to the nonlocal phase modulation as shown in Fig.3(b). However, at $0^{\circ}$ phase difference, the \textit{X} and \textit{P} quadratures of the probe and conjugate beams are coupled, and positive correlations can be seen. The double diagonal structure in the covariance matrix corresponds to the frequency sidebands introduced by the EOMs and demonstrates the mulitimode nature of the phase-modulated joint field quadratures. A single EOM, driven at twice the modulation index produces similar correlations to those produced by two in-phase EOMs. The \textit{XP} measurements are phase sensitive. Not only the relative phase of the EOMs is important, but the phase of the data windows with respect to the EOM drive also matters. In all cases, while changing the driving phase of the conjugate EOM, we kept the phase of the probe EOM fixed at $0^{\circ}$. 

In all results presented so far, both EOMs were placed in the path of the twin beams. This configuration is of interest for quantum information processing applications, such as the production of cluster states or quantum key distribution. An equivalent measurement can be made with the EOM in a local oscillator beam. This does not create an entangled state, but the measurement result is the same as if it did. Recent proposals pointed out that multi-mode homodyne detection can realize compact Gaussian quantum computation by selecting appropriate LO measurement choices~\cite{ferrini2013compact,Cai2015, ferrini2016direct}, which includes digital post-processing. In Fig.~4(b), we show the results obtained when both EOMs are placed in the LO paths. It is clear that the effect of the EOMs on the measured beam correlations is the same as the one observed with the EOMs placed in the beams, except for a change of sign in the correlations. Placing one EOM in the probe beam and the other EOM in the LO of the conjugate beam causes a different effect on the measured correlations, as shown in Fig.~4(c). In this case, the EOMs cancel each other when they are in phase and couple the \textit{X} and \textit{P} quadratures when they are out of phase.  

One can view the arrangement in Fig.~1 as a “truncated” version of the SU(1,1) interferometer~\cite{anderson2017} that requires homodyne detection to read out the phase. This configuration can be seen to be a pair of interferometers, each comprised of a (very noisy) signal beam, plus a LO beam. These interferometers will, however, have quantum-correlated signals, and any phase shift written onto one of the beams can be detected at a sub-shot-noise level in the difference signals. When one views the interferometers independently, it is clear that it does not matter if the phase shift is written onto the “signal” beam, or onto the LO beam – either way, the homodyne output contains the signal.  Because of the geometry, however, a similar phase shift written on the LO will appear as a phase shift of the opposite sign to one written onto the signal beam. If the signal and the local oscillator have the same phase shifts, the detector will not see it.

\begin{figure}[htbp]
\centering
\includegraphics[width=8.5cm]{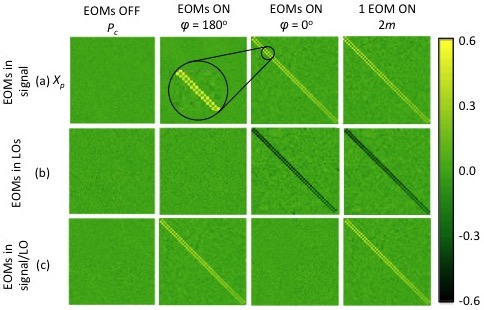}
\caption{Measured $X_p P_c$ covariance blocks for (a) both EOMs in the twin beams, (b) both EOMs in the local oscillator beams and (c) one EOM in the probe beam and the other EOM in conjugate local oscillator beam. When both EOMs are on, their modulation index is $m = 0.1\pi$. The double diagonal structure of the correlations (magnefied inset) corresponds to the first-order frequency sidebands due to the periodic modulation of the beams by the EOMs. For each square, the horizontal and vertical axes correspond to 200~kHz frequency bins spanning the range 200~kHz to 10~MHz~\cite{Suppl2023}.}
\label{fig5}
\end{figure}

In conclusion, we have studied the effects of electro-optical phase modulation on two-mode squeezing of multi-frequency-mode, continuous variable twin beams. The probe and conjugate modulations interfere nonlocally to modify the beam correlations, which are controlled by adjusting the relative driving phase of the modulators. We found that the modulators acted cumulatively to determine the effective modulation index.  We believe that our setup is a potential platform for further experimental studies on cluster state generation, quantum erasing, and quantum sensing. The ability to manipulate twin beam correlations via nonlocal phase modulation has important implications for those fields. For example, positioning the EOM in the local oscillator should allow the implementation of compressed sensing for quantum system characterization, thereby measuring the appropriate frequency mode combinations, rather than mixing the modes directly. Positioning the EOMs in the local oscillators can also bring an experimental advantage, since it avoids introducing additional losses in the signal beams, allowing for larger squeezing signals. A recent proposal for generating hypercubic cluster states~\cite{zhu2021} suggests using an EOM to couple different frequency qumodes of two-mode entangled beams and is a natural next step for the present experiments. We have found that in generating entangled states, it is completely equivalent to use a single EOM in one of the twin beams or to have an EOM in each beam. A complex waveform may be required, but the nonlocal nature of the modulation allows all of the modulation to take place in one beam. Because the bandwidth of our squeezed light is limited, we are able to digitize the measurements across the entire spectrum. This would allow one to implement the direct approach to measurement-based computing suggested in~\cite{ferrini2013compact, Cai2015, ferrini2016direct}.

\begin{acknowledgments}This work was supported by the Air Force Office of Scientific Research (FA9550- 16-1-0423). Lu\'{i}s E.~E.~de Araujo acknowledges the financial support of grant \#2019/24743-9, S\~{a}o Paulo Research Foundation (FAPESP). We acknowledge Alessandro Restelli for the help with the lock circuits. This research was performed while Matthew DiMario held a National Research Council Research Associateship at NIST. 
\end{acknowledgments}


\begin{thebibliography}{37}
\bibitem [{noi()}]{noise}%
  \BibitemOpen
  \href@noop {} {\ }\bibinfo {note} {To describe the entanglement between two
  fields, such as $\hat{A}_\text{p}$ and $\hat{A}_\text{c}$, we define the
  joint quadrature operators: $\hat{X}_{\pm} = (\hat{X}_\text{p} \pm
  \hat{X}_\text{c})$ and $\hat{P}_{\pm} = (\hat{P}_\text{p} \pm
  \hat{P}_\text{c})$, where $\hat{X}_\text{p,c} = (\hat{A}_\text{p,c} +
  \hat{A}_\text{p,c}^{\dagger})/\sqrt{2}$ and $\hat{P}_\text{p,c} =-i
  (\hat{A}_\text{p,c} - \hat{A}_\text{p,c}^{\dagger})/\sqrt{2}$ are the usual
  amplitude and phase quadratures, respectively, of the fields.}\BibitemShut
  {Stop}%
\end{thebibliography}

%

\end{document}